\documentclass[11pt,a4paper]{amsart}

\usepackage{cite}

\usepackage[latin1]{inputenc}
\usepackage[english]{babel}
\usepackage{amsmath}

\usepackage{amssymb, mathabx}
\usepackage{graphicx}
\usepackage{braket}
\usepackage[pdftex,plainpages=false,colorlinks,hyperindex,bookmarksopen,linkcolor=red,citecolor=blue,urlcolor=blue]{hyperref}
\usepackage{caption}
\usepackage{subcaption}

\usepackage{mathrsfs}

\usepackage{epstopdf}

\usepackage{braket}

\usepackage[hmargin=3cm,vmargin={3.5cm,4cm}]{geometry}

\theoremstyle{theorem}

\theoremstyle{definition}                                 %stile corsivo
\newtheorem{claim}{Claim}                       %definizione ambiente teorema
\newtheorem{proposition}[claim]{Proposition}      %definizione ambiente proposizione

\theoremstyle{definition}                           %stile roman
\newtheorem{definition}{Definition}                   %definizione ambiente definizione

\theoremstyle{remark}                             %stile per osservazioni
\usepackage{color}

\usepackage{mathtools,slashed}

\newcommand{\be}{\begin{eqnarray}}
\newcommand{\ee}{\end{eqnarray}}

\newcommand{\R}{\mathbb{R}}  %%%%% \R = \mathbb{R}.
\def\eg{{\it e.g. }} 
\def\ie{{\it i.e. }}

\newcommand{\ceil}[1]{\lceil #1 \rceil}

\def\eg{{\it e.g.}\ }
\def\ie{{\it i.e.}\ }

\numberwithin{equation}{section}

\allowdisplaybreaks

\begin{document}
\title{Modeling biological systems with an improved fractional Gompertz law}
	
	   \author{Luigi Frunzo$^1$}
		\address{${}^1$ University of Naples ``Federico II'', Department of Mathematics and
		Applications, via Cintia 1, 80126, Naples, ITALY.}
		\email{luigi.frunzo@unina.it}
		
		\author{Roberto Garra$^2$}
		\address{${}^2$ Department of Statistical Sciences, 
		 Sapienza, University of Rome. P.le Aldo Moro, 5,
		 Rome, ITALY} 	
		\email{roberto.garra@sbai.uniroma1.it}
	
	    \author{Andrea Giusti$^3$}
		\address{${}^3$ Department of Physics $\&$ Astronomy, University of 	
    	    Bologna and INFN. Via Irnerio 46, Bologna, ITALY and 
	    	 Arnold Sommerfeld Center, Ludwig-Maximilians-Universit\"at, 
	    	 Theresienstra{\ss}e~37, 80333 M\"unchen, GERMANY.}
		\email{andrea.giusti@bo.infn.it}
	
   		\author{Vincenzo Luongo$^4$}
    	    \address{${}^4$ University of Naples ``Federico II'', Department of Mathematics and
		Applications,  via Cintia 1, 80126, Naples, ITALY.}
		\email{vincenzo.luongo@unina.it}
 
	
	\thanks{In: \textbf{Comm.~Nonlinear~Sci.~Numer.~Simulat.~(2019)}, \textbf{DOI}: \href{https://doi.org/10.1016/j.cnsns.2019.03.024}{10.1016/j.cnsns.2019.03.024}}
	
    \date  {\today}%%{January 2016}

\begin{abstract}
The aim of this paper is to provide a fractional generalization of the Gompertz law via a Caputo-like definition of fractional derivative of a function with respect to another function. In particular, we observe that the model presented appears to be substantially different from the other attempt of fractional modifications of this model, since the fractional nature is carried along by the general solution even in its asymptotic behavior for long times. We then validate the presented model by employing it as a reference frame to model three biological systems of peculiar interest for biophysics and environmental engineering, namely: dark fermentation, photofermentation and microalgae biomass growth.
\end{abstract}

    \maketitle

  \section{Introduction}
     
     The Gompertz curve was first introduced in \cite{gompertz} as an empirical model for describing the human age distribution in a given community,  within the analysis of mortality tables. Then, this distribution has found many applications in biophysics, with particular regard to the mathematical modeling of tumor growth.
     
     The Gompertz law, from a mathematical perspective, is a Malthusian-type growth model with a time-dependent exponentially decreasing rate. The stochastic roots of this model have been carefully addressed by De Lauro et al. in \cite{delauro} and an interesting generalization of this model has been recently presented by Di Crescenzo and Spina in \cite{dicre}.
     
     Applications of fractional calculus \cite{Giusti-Arxiv, Mainardi-Gorenflo-1997, Mainardi-1997, epirozzi} to counting processes and population modeling has gained increasing attention over the last few years. This can clearly be inferred from the extent of the current literature devoted to fractional Poisson processes \cite{enzo,KY,laskin,fra}, fractional birth-death processes \cite{O1,O2} and fractional Malthusian models \cite{alm1}.
      Further, certain diffusion processes appear to be accurately modeled in terms of fractional kinetic equations \cite{ralf}. Such dynamics are known to play a key role in several mechanisms involved in the life and development of microorganisms. Indeed, as discussed in \cite{ToledoHernandez}, it is reasonable to conceive that the physicochemical nature of some relevant biological processes, particularly those involved in cell growth and enzymatic reactions, will likely show the emergence of memory effects.
		
     Aside from some interesting modifications involving techniques of ordinary calculus (see \eg \cite{PLOS}), a first attempt to consider a generalization of the Gompertz law, by means of fractional calculus tools, has been proposed by Bolton et al. in \cite{bolton}. Besides, very recently, Almeida and collaborators \cite{almeida,alm1,alm2} have shown the peculiar relevance, for physical applications, of the notion of fractional derivative of a function with respect to another function. 
     
     Inspired by these investigations, in this paper we suggest a new generalization of the Gompertz model based on the mathematical approach developed by Almeida in \cite{almeida}. 
     
     We compare this new approach with the previous one developed by Bolton et al. \cite{bolton} and characterize its main features. Then, we test the proposed model on three distinct biological systems that are of particular interest for applications in environmental engineering. In this regard, let us stress that the key point of this paper is the  fractional modification of the Gompertz law via a Caputo-like definition of fractional derivative of a function with respect to another function. The experimental validation, even though it involves some original and unpublished data that might very well be appropriately described in terms of traditional tools, is therefore regarded as a test of the general features of the proposed model.

     \section{Fractional Gompertzian-type models: previous and new results} \label{sec-generaltheory}
	The seed behind our idea for a generalization of the Gompertz law via fractional calculus techniques is inspired by a recent study by Bolton et al. \cite{bolton}. Indeed, in \cite{bolton} the authors tested their proposal for a fractional improvement of the aforementioned empirical law by comparing their theoretical model with an experimental dataset concerning the volume growth of the Rhabdomyosarcoma tumor in mice. 
	
	Let us first recall the key ideas behind the original formulations of both the Gompertz law and its modification proposed by Bloton and collaborators. 
	
	The original formulation of Gompertz's model arises from the following evolution equation,
	\begin{equation}\label{a}
     \frac{1}{V}\frac{dV}{dt} = \alpha- \beta \ln \left( \frac{V}{V_0} \right) \, ,
     \end{equation}	
where $V = V(t)$ is the volume of a tumor at time $t$, $V_0 = V(0)$, while $\alpha$ and $\beta$ are the so called \textit{kinetic parameters} of the tumor (see \eg \cite{whelton}). Denoting by $y = \ln(V/V_0)$, Eq.~\eqref{a} reduces to
     \begin{equation}\label{b}
     \frac{dy}{dt}= \alpha-\beta y 
     \end{equation} 
    whose solution is given by $y(t)= \frac{\alpha}{\beta}(1-e^{-\beta t})$. 
 
 Then, going back to the physical variable $V(t)$, the \textit{Gompertz law} is obtained in its well-known form 
    \begin{equation}
   V(t) = V_0 \, \exp \left[ \frac{\alpha}{\beta} \, \left(1-e^{-\beta t} \right) \right] \, .
    \end{equation}
	
	It is also worth remarking that the Gompertz law is also recovered as a simple Malthusian model (see \cite{dicre}), namely
	\begin{equation} \label{malt1}
    \frac{dN}{dt} = \lambda(t) \, N \, ,
    \end{equation}
    where $N = N(t)$ now represents the population size at time $t$, with an exponentially time-decreasing rate
    \begin{equation} \label{malt2}
    \lambda(t) = \alpha e^{-\beta t} \, .
    \end{equation}
	
	  The fractional generalization considered by Bolton et al. in \cite{bolton} is obtained by replacing the first order time derivative in \eqref{b} with a Caputo fractional derivative of order $\mu \in (0, 2)$. According to their analysis, it seems that the physically meaningful range for the fractional order parameter $\mu$ is $\mu \in (0,1]$ (in particular, the best agreement between the model and the data is obtained for $\mu = 0.68$). 	This result can be further justified by the fact that the solution of this improved model involves the Mittag-Leffler function that has an oscillating behavior for $\mu >1$. Indeed, the general solution of the evolution equation discussed in \cite{bolton} is simply given by
	\begin{equation}\label{bol}
    V_{\mu}(t) = V_0 \, \exp \left\{ \frac{\alpha}{\beta} \left[1-E_{\mu}(-\beta t^\mu)\right]\right\},
    \end{equation}
    where 
    \begin{equation}
   E_{\mu}(-\beta t^\mu) = \sum_{k=0}^\infty \frac{(-\beta t^\mu)^k}{\Gamma(\mu k+1)}, 
    \end{equation}
    is the well-known Mittag-Leffler function, that plays a key-role in the theory of fractional differential equations (see \eg the recent monograph \cite{ml}, and \cite{Garrappa1, Garrappa2} for a precise discussion on its numerical evaluation).   
    
    \begin{figure}[h!]
    \centering
    \includegraphics[width=9cm]{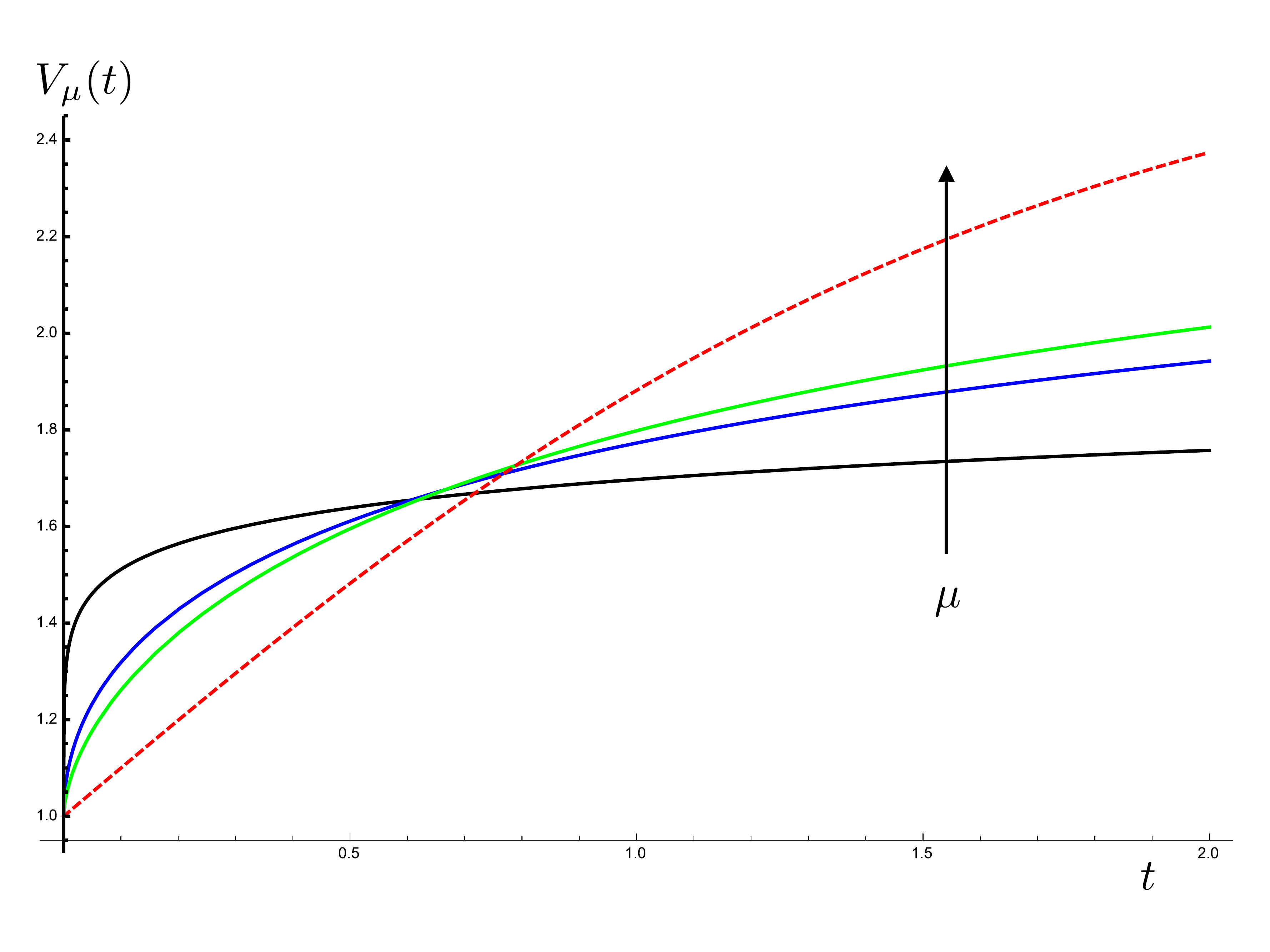}
    \caption{Plot of $V_\mu (t)$ in Eq.~\eqref{bol} with $\alpha = \beta = V_0 = 1$ and $\mu = 0.2, \, 0.5, \, 0.6, \, 1.0$ (\ie the dashed line represents the ordinary Gompertz law).}
    \end{figure}
	  
	In this paper, we provide an alternative analysis of Gompertz's evolution equation based on the application of fractional Caputo derivatives with respect to another function. This idea is inspired by the recent studies by Almeida and coauthors \cite{almeida, alm1, alm2}, where this new approach was applied to the study of population growth processes.  
	  
	  Our approach differs substantially from the one considered by Bolton et al. in \cite{bolton} due to the different implementation of the fractionalization. Indeed, if we consider the model equation for the Malthusian model that leads to a Gompertz growth, namely Eq.~\eqref{malt1} with the time dependent rate in Eq.~\eqref{malt2}, that reads
	  \be \label{Gompertz-Constitutive}
	  e^{\beta \, t} \, \frac{d N}{dt} =\alpha \, N \, ,
	  \ee
we can simply perform a fractionalization of the latter by introducing the fractional derivative of $N(t)$ with respect to $e^{\beta t}$.

	Hence, the new model equation becomes
	    \begin{equation}\label{ngr}
       {}^C \left(e^{\beta t}\frac{d}{dt}\right)^\mu  N_\mu(t) =\alpha N_\mu(t), \quad \mu\in(0,1], 
     \end{equation}
     with the initial condition $N_\mu (0) = N_0 > 0$, where
      \begin{equation} \label{our-frac}
      {}^C \left(e^{\beta t}\frac{d}{dt}\right)^\mu  N_\mu(t)  =\frac{1}{\Gamma(1-\mu)}\int_0^t \left(\frac{e^{-\beta \tau} - e^{-\beta t}}{\beta} \right) ^{- \mu} \, \frac{d N_\mu}{d \tau} \, d \tau \, ,
     \end{equation}
     which, as mentioned above, is a particular case of the Caputo-type fractional derivative of a function with respect to another function (see \eg \cite{almeida} and \ref{App-A} for further details). It is also worth remarking that this approach has been used to treat a Dodson-type fractional diffusion equation in \cite{noi}.
     
     It is easy to recognize that Eq.~\eqref{ngr} is an eigenvalue problem for the Caputo-type fractional operator defined in Eq.~\eqref{our-frac}, whose solution is given by
     \begin{equation}\label{new}
     N_\mu(t) = N_0 \, E_\mu \left[\frac{\alpha}{\beta^\mu}\left(1-e^{-\beta t}\right)^\mu\right] \, .
     \end{equation}
     In \ref{App-A} we also provide the solution for a general eigenvalue problem involving a Caputo-type fractional derivative of a function with respect to another function and we explain how the general result applies to the case at hand.
     
\begin{figure}[h!]
   \centering
    \includegraphics[width=14cm]{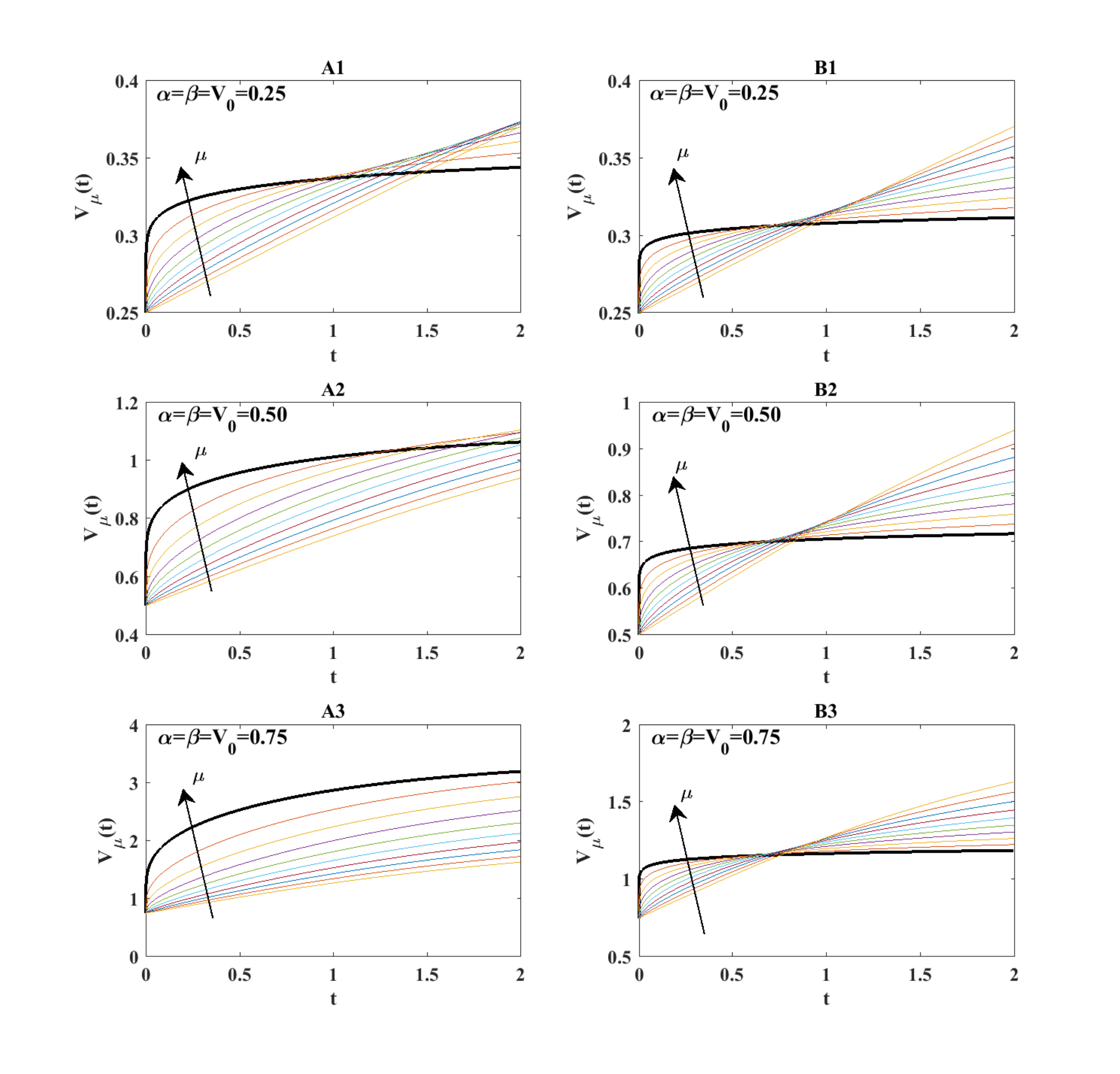}
    \caption{Comparison, for different values of the parameter $\mu$ $(0.1 \leq \mu \leq 1)$, between the generalized Gompertz law \eqref{ngr} presented here (left: $A1$, $A2$, $A3$) and the model by Bolton et al. \eqref{bol} (right: $B1$, $B2$, $B3$). The arrows show how the solutions change as we increase the parameter $\mu$. Further, the thicker solid line in all panels represents the case $\mu = 1$ (ordinary Gompertz).}
    \label{comp-us-bol}
    \end{figure}

	Although for both Eq.~\eqref{bol} and Eq.~\eqref{new} the classical Gompertz law is recovered for $\mu =1$, the main features of the two models clearly differ, as it can be seen in \figurename~\ref{comp-us-bol}. Indeed, in the first case we have the composition of a fastly increasing function with the slowly decaying one, the latter being the Mittag-Leffler function, whereas in the second case the situation is reversed. Indeed, we have that
     \begin{equation} \label{eq-asymptotic-new}
     \lim_{t\rightarrow +\infty} N_\mu (t) = N_0 \, E_{\mu}\left(\frac{\alpha}{\beta^\mu}\right),
     \end{equation}
     while 
     \begin{equation} \label{eq-asymptotic-bolton}
     \lim_{t\rightarrow +\infty} V_\mu(t) = \lim_{t\rightarrow +\infty} V_1(t) = V_0 \, \exp{(\alpha/\beta)} \, .
     \end{equation}
     This implies that, asymptotically, the model by Bolton et al. has a classical Gompertz behavior, while this is not the case for the model introduced in this work. The substantial difference between these two models is further highlighted by the illuminating plots in \figurename~\ref{comp-us-bol}.
      
     From a physical perspective, the generalization of the Gompertz law introduced in Eq.~\eqref{ngr} has the property of embedding the deterministic time-change (due to the variable rate) 
		in the definition of the fractional operator. The result of this modification is then the emergence of memory effects in the population dynamics. 
		Indeed, this can clearly be inferred from Eq.~\eqref{new} by the peculiar fractional-power law behavior and the distinctive departure 
		from the typical exponential picture for long times. Besides, the non-local nature of the operator in Eq.~\eqref{our-frac} is also self-evident in light of Tarasov's criteria \cite{Tarasov}.

		The use of fractional calculus to model bacterial population dynamics is, in our opinion, particularly interesting as it allows us to take into account phenomena like ``biomass adaptation'' or ``lag-phase'', that turn out to be particularly important in batch experiments like those presented in this work. Along this line, it is worth to stress that several biological studies have shown that bacteria tend to change their features adapting to different environmental conditions. Of course, there are several ways to actively model this adaptation, from both physical and biological perspectives, yet a common trait of these schemes, particularly when dealing with adaptation strategies of microbial biomass, is the need to take into account the whole history of the microbial population used for ``in-vitro'' experiments. Furthermore, the dynamics of most reactive biological systems is modeled by means of ordinary differential equations, which however completely neglect any spatial effect. This is usually justified by the assumption that bacterial cells live suspended in the bulk liquid. Nonetheless, it was recently observed that bacteria tend to organize themselves in biofilm form \cite{Flemming}, which are aggregates of cells enclosed in a self-produced matrix \cite{Mattei2018, DAC2019}. Then diffusion phenomena at micro-scale turn out to be of paramount importance for such systems. In light of this, it was shown that fractional models can appropriately account for these anomalous kinetics \cite{ToledoHernandez}.
     
\section{Applications and numerical solutions}\label{n5}

	In the following, we present an application of the model presented in Section~\ref{sec-generaltheory} to three different biological systems of major interest for environmental engineering. The first two cases deal with the biological production of gas (biogas) catalysed by two different fermentation processes. The third application, instead, concerns %an analysis of
 the growth of a selected microalgae biomass in non-limiting nutrient conditions.

	Over the last decade, many studies on biogas production through anaerobic fermentative systems have been reported, see \eg \cite{ghimire2015review,de2013review}. Due to the production of a high energy content gaseous product, such as hydrogen, accompanied solely by water during its combustion, Dark Fermentation (DF) is widely regarded as one of the most promising anaerobic biological process for engineering purposes \cite{DeGioannis2014}. This process is usually performed by multispecies anaerobic consortia, which provide the conversion of organic substrates, \eg the organic
fraction of municipal solid wastes, in a biologically derived fuel that can be used to produce energy. The wide range of organic (solid-liquid) waste compounds that can act as feedstock for the DF has led this process to become one of the key technique for renewable energy production in both laboratories and pilot scale reactors.

	Similar results can be achieved by employing Photofermentative bacteria, such as \textit{Rodobacter Sphaeroides} (Second application). These species are able to catalyze the conversion of different organic compounds by using light as a supplementary energy source \cite{ghimire2016concomitant, trchounian2017improving}. 
	In this case, the organic soluble compounds, which are usually derived from a clarified wastewater, are involved in biochemical reactions that lead to the accumulation of poly-hydroxybutyrate (PHB) and the contextual production of hydrogen. PHBs are value-added biochemical compounds as they can be easily extracted from the bacterial cells and processed to produce bio-plastic materials \cite{luongo2017photofermentative}. Besides, the hydrogen produced in these process can be used as a renewable energy source, as it ensure a high energy recovery from organic wastes.

	Finally, the third application of the model introduced in Section~\ref{sec-generaltheory} deals with the description of the microalgae biomass growth. These unicellular micro-organisms can be used in the wastewater treatment field for the removal of nutrients, as an alternative treatment to the traditional, energy demanding, aerobic systems \cite{iasimone2017}. Microalgae can assimilate a large amount of inorganic nitrogen and phosphorous in autotrophic conditions by employing carbon dioxide as inorganic carbon source and mitigating greenhouse gas emissions with the contextual production of oxygen. Moreover, these micro-organisms, depending on the conditions at hand, can shift their metabolism from autotrophic to heterotrophic or, even, assimilate organic compounds to produce hydrogen \cite{extra}. In addition, many microalgae species are able to accumulate value-added compounds during their growth, usually due to the high protein content of their cell membrane, which can be later extracted or used for biofuel production.

	The complete description of the lab-scale procedures adopted for the experimental data acquisition have been reported in the Supplementary Materials (\ref{sup}), whereas the model application and validation is reported in the following Section.

\subsection{Model validation}\label{n5.4}

\begin{figure}[h!]
\begin{center}
\includegraphics[width=1.05 \textwidth]{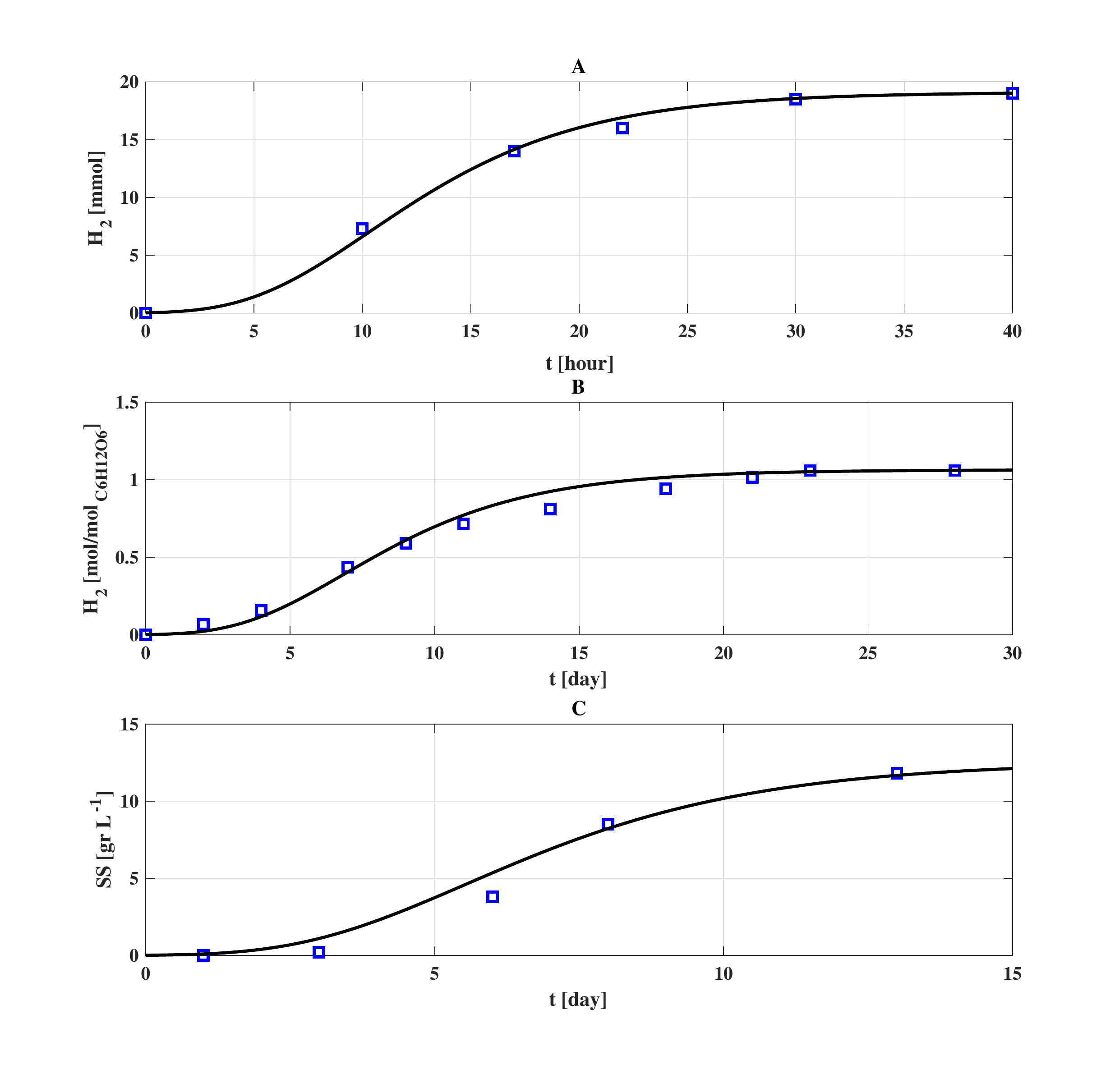}
\caption{Comparison between experimental data and numerical simulations for the three experimental cases: (A) Dark Fermentation, (B) Photo fermentation, (C) Algae.}
 \label{f1.4} 
 \end{center}
 \end{figure}

The modified Gompertz law proposed in this work has been applied to the three experimental cases described above. The calibrated parameters used in numerical simulation are reported in Table~\ref{t4.1}, whereas a comparison between experimental data and numerical simulations for each case is reported in \figurename~\ref{f1.4}.

\begin{table}[h!]
\caption{Model parameters} \label{t4.1}
\begin{small}
 \begin{center}
 %\begin{tabular}{lc{3cm}lccccc}
 \begin{tabular}{lclc}
 \hline\hline
{\textbf{Parameter}} & {\textbf{A}} & {\textbf{B}} &  \textbf{C} \\
 \hline
 $\alpha$            &  $1.0965$      &  $1.5200$       & $2.1350$   \\
 $\beta$             &  $0.180$       &  $0.275$        & $0.350$    \\
 $V_0$               &  $0.020$       &  $0.001$        & $0.010$    \\
 $\mu$               &  $0.64$        &  $0.70$         & $0.85$     \\
 \hline\hline
 \end{tabular}
 \end{center}
 \end{small}
 \end{table}
 
	The quantitative comparison between observed and predicted data has been carried out via three standard methods for model validation \cite{Janssen, frunzo}, namely the normalized root mean square error (nRMSE), the index of agreement (IoA), and the modeling efficiency (ME). Specifically,
     \begin{equation} \label{eqnRMSE}
     {\rm nRMSE}= \frac{1}{\overline{O}} \sqrt{\frac{\sum_{i=1}^{N}\left(P_i-O_i\right)^{2}}{N}} \, , 
     \end{equation}
	\begin{equation} \label{eqIoA}
     {\rm IoA} 
     = 
     1 - \frac{\sum_{i=1}^{N}\left(P_i-O_i\right)^{2}}{\sum_{i=1}^{N}\left(\left|P'_i\right|+\left|O'_i\right|\right)^{2}}
     \, ,
     \end{equation}
     \begin{equation} \label{eqME}
     {\rm ME} 
     = 
     \frac{\sum_{i=1}^{N}\left(O_i-\overline{O}\right)^{2} 
     - \sum_{i=1}^{N}\left(P_i-O_i\right)^{2}}{\sum_{i=1}^{N}\left(O_i-\overline{O}\right)^{2}};
     \end{equation}
where $P_i$ and $O_i$ denote the predicted and observed values, $i=1,\dots,N$, $N$ is the number of observed/predicted values, $\overline{O}$ is the mean of observed data, and $P'_i=P_i-\overline{O}$; $O'_i=O_i-\overline{O}$.

The results of the validation procedure are then depicted in \figurename~\ref{f1.4}. Besides, a comparison between 
experimental data and numerical simulations is also summarized in Table~\ref{t4.2}.

\begin{table}[h!]
\caption{Results of model validation} \label{t4.2}
\begin{small}
 \begin{center}
 %\begin{tabular}{lc{3cm}lccccc}
 \begin{tabular}{lclc}
 \hline\hline
{\textbf{Experiment}} & {\textbf{nRMSE}} & {\textbf{IoA}} &  \textbf{ME} \\
 \hline
 Dark Fermentation    &  $0.0374$        &  $0.9988$      & $0.9952$    \\
 Photofermentation    &  $0.0792$        &  $0.9961$      & $0.9835$    \\
 Microalgae           &  $0.1672$        &  $0.9918$      & $0.9693$    \\
 
 \hline\hline
 \end{tabular}
 \end{center}
 \end{small}
 \end{table}

 \section{Conclusion}\label{n6}
	In this paper we introduced a mathematical model that generalizes the well-known Gompertz law of population dynamics by including fractional features in its constitutive (ordinary) differential equation. Specifically, this procedure was performed by replacing the left-hand side in Eq.~\eqref{Gompertz-Constitutive} with the Caputo-type derivative of $N(t)$ with respect to $\exp(- \beta \, t)$. The resulting model is then governed by a purely fractional differential equation whose solution of the corresponding initial value problem \eqref{new} shows that the fractional nature of the described theoretical set-up is carried along for the whole history of the process. Indeed, one of the key differences between the behavior of \eqref{new} and the model by Bolton et al \eqref{bol} is that for the latter the contribution of the fractional feature get slowly turned off at late time, ultimately stabilizing the evolution to an asymptotic value \eqref{eq-asymptotic-bolton} which does not depend on the fractional parameter $\mu$. Whereas, the new growth law featured in \eqref{new} is still strongly dependent on the fractionalization procedure as shown in \eqref{eq-asymptotic-new}, that underlines an enhanced capability of this model to fit a broader class of experimental behaviors.  
	
	In order to further strengthen our arguments, we have tested the performance of the presented model in three very peculiar and distinct frameworks, namely: two experiments regarding the biogas production catalyzed by either DF (\ref{n5.1}) or PF (\ref{n5.2}), and one experiment featuring microalgae biomass growth (\ref{n5.3}). The results concerning the calibration and numerical simulations are reported in Section~\ref{n5.4}, together with some illuminating plots, see \figurename~\ref{f1.4}.
     
     Lastly, a comment is in order concerning the general reasoning behind the very significance of the whole fractionalization process. The underlying motivation for this work was to analyze what kind of information could be retrieved by the introduction of the newly developed non-local operator that, as stressed above, clearly improves on previous attempts at generalizing the Gompertz law. Then, we decided to test the effectiveness of these modified paradigm upon experimental data which are usually tackled with the standard Gompertz law. Of course, it is well known that not all biological systems follow the usual Gompertz law, so this is per se a valid justification for looking for evolution equations whose solution might even depart sensibly form the latter. However, the typical issue that is commonly raised when dealing with fractional operators is connected with their vindication starting from first principles. This is usually a very difficult matter to address since, most of the time, it requires a case-dependent bottom-up construction of the theory (see \eg \cite{ralf, gianni-1, AG-FM_MECC16, gianni-2, IC-AG-FM-ZAMP} and references therein). Hence, we leave this matter regarding the hereby presented modified Gompertz law for a future study.
     
 \appendix    
     
      \section{Preliminaries about Caputo fractional derivative of a function with respect to another function} \label{App-A}
          
          Fractional derivatives of a function with respect to another function have been part of the fractional calculus literature at least since the publication of the classical monograph by Kilbas et al. \cite{kilbas} (see Section 2.5) and appear to have found a new life tanks to some recent works by Almeida \cite{almeida}.
          
          In this Appendix, we briefly recall the main definitions and properties of these operators. Nonetheless, this section cannot represent an exhaustive description of these peculiar operators, hence we invite the interested reader to refer to \cite{almeida} for a more complete discussion. We also observe that this approach has found some interesting applications in \cite{alm1, alm2, noi, noi-2, orega}.
          
          \begin{definition} \label{Def-I}
          Let $\nu>0$, $I = (a,b)$ be an interval such that $-\infty \leq a < b \leq +\infty$, $g(t) \in L_1 (I)$ and $f(t)\in C^1(I)$ strictly increasing function for all $t\in I$.
	Then, the fractional integral of a function $g(t)$ with respect to another function $f(t)$ is given by 
\begin{equation} \label{eq-def-I}
I^{\nu,f}_{a+}g(t):=\frac{1}{\Gamma(\nu)}\int_a^t f'(\tau)
[f(t)-f(\tau)]^{\nu-1}g(\tau)d\tau.
\end{equation}
          \end{definition}
         
	By inspection one can immediately notice that for $f(t) = t^{\alpha/\beta}$ we recover the Erd\'elyi-Kober fractional integral (see \eg \cite{gia1, noi-3} for some physical applications), whereas if we set $f(t)= \ln t$, this integral can be recasted in the Hadamard fractional integral. Finally, for $f(t)= t$ one trivially recovers the Riemann-Liouville fractional integral.
          
     Now, given Definition~\ref{Def-I} one can introduce a corresponding regularized left-inverse operator of \label{eq-def-I} given by,
  
         \begin{definition} \label{def-der}
        Let $\nu>0$, $n \equiv \ceil{\nu}$, $I = (a,b)$ be an interval such that $-\infty \leq a < b \leq +\infty$, $g(t) \in C^n (I)$ and $f(t)\in C^1(I)$ strictly increasing function for all $t\in I$. 
        Then, the Caputo derivative of the function $g(t)$ with respect to a function $f(t)$ is given by
           \begin{equation}\label{eq-def-der}
           {}^C \left(\frac{1}{f'(t)}\frac{d}{dt}\right)^\nu g(t) :=I_{a+}^{n-\nu,f}\left(\frac{1}{f'(t)}\frac{d}{dt}\right)^n g(t) \, .
               \end{equation} 
          \end{definition}

	For the latter it is easy to infer that, 
          
          \begin{proposition}
          Let $\nu>0$, $n \equiv \ceil{\nu}$, $I = (a,b)$ be an interval such that $-\infty \leq a < b \leq +\infty$ and $f(t)\in C^1(I)$ strictly increasing function for all $t\in I$. If $g(t) =[f(t)-f(a+)]^{\beta-1}$ with $\beta>1$, then
          \be 
          {}^C \left(\frac{1}{f'(t)}\frac{d}{dt}\right)^\nu  g(t) = \frac{\Gamma(\beta)}{\Gamma(\beta-\nu)} [f(t)-f(a+)]^{\beta-\nu-1} \, .
          \ee
          \end{proposition}
          
 	\begin{proposition} \label{prop-2}
          Let $\nu>0$, $n \equiv \ceil{\nu}$, $I = (a,b)$ be an interval such that $-\infty \leq a < b \leq +\infty$ and $f(t)\in C^1(I)$ strictly increasing function for all $t\in I$. If $g(t) =E_\nu [\lambda \, (f(t)-f(a+))^\nu]$ with $\lambda \in \R$, then
          \be 
          {}^C \left(\frac{1}{f'(t)}\frac{d}{dt}\right)^\nu  g(t) = \lambda \, g(t) \, .
          \ee
          \end{proposition}             
          
          The operator appearing in the generalized Gompertz equation \eqref{ngr} is obtained from the general definition \eqref{eq-def-der} by setting $a = 0^+$ and choosing $f(t)$ such that $f'(t) =e^{-\beta t}$. Hence, from \ref{prop-2} we have that the eigenfunction of the operator \eqref{our-frac} is given by 
          \begin{equation}
          N_\mu(t) = E_\mu\left[\frac{\alpha}{\beta^\mu}(1-e^{-\beta t})\right] \, .
          \end{equation}		

\section{Supplementary Materials} \label{sup}

	In this Section we report a detailed description of the three experimental setups for the model validation discussed in the manuscript.

\subsection{Dark Fermentation experiments}\label{n5.1} 

 DF experiments have been inoculated with a pre-treated full-scale anaerobic digestion sludge. 
 The thermal pre-treatment leads to the inhibition of methanogenic bacteria with consequent
 hydrogen accumulation during the fermentation process. 
 
 \begin{figure}[h]
 \begin{center}
 \includegraphics[width=.6\textwidth]{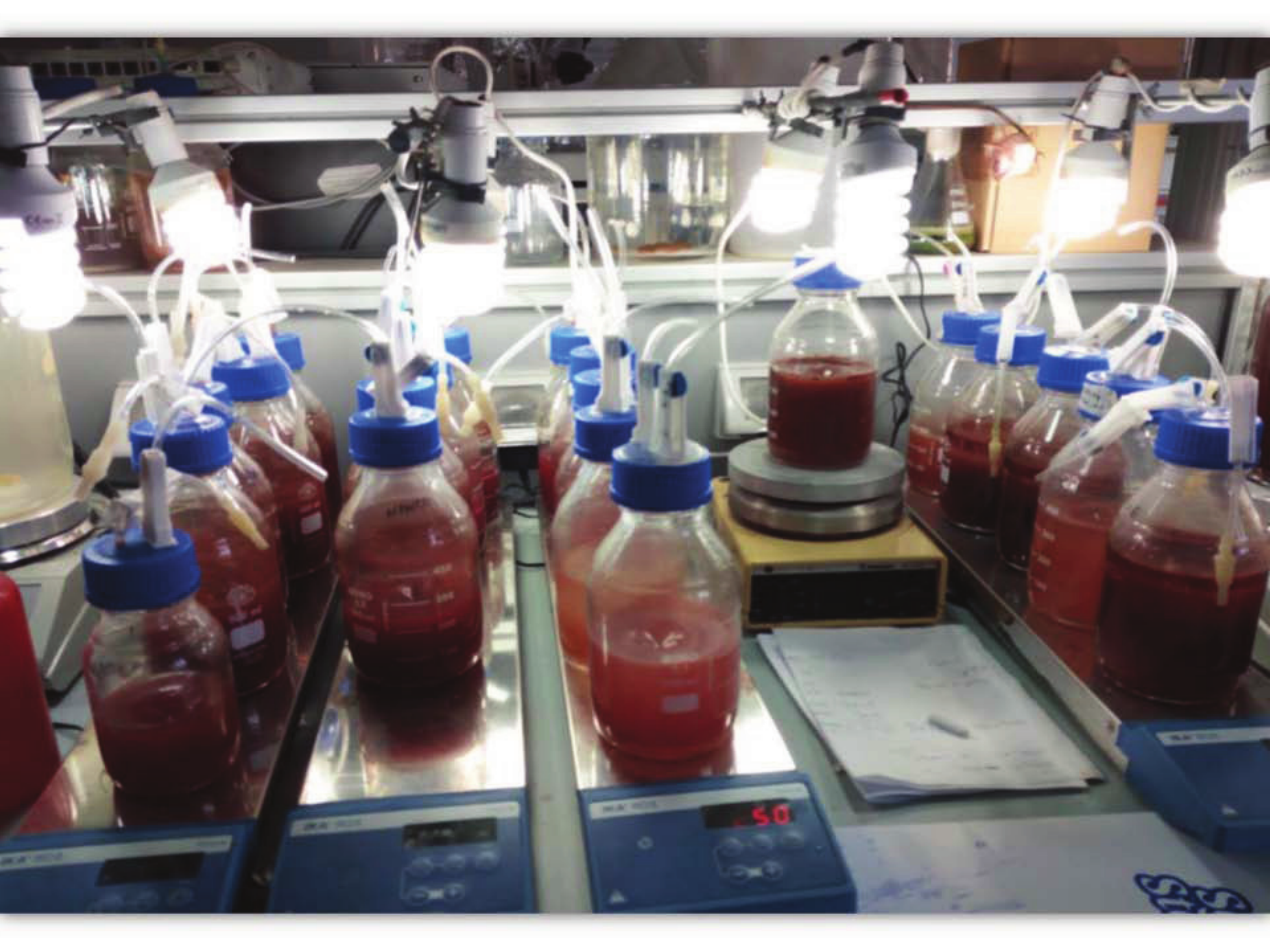}
 \caption{Photofermentative experiment}
 \label{f1.1}
 \end{center}        
 \end{figure}

 According to \cite{ghimire2015review}, the
 collected sludge was retained in a lab-scale oven for one hour at $105 \, {}^{\circ} C$. The anaerobic sludge was characterized in terms of Total Solids (TS), Volatile Solids (VS) and pH, according to the standard methods \cite{federation2005standard}. The results showed TS and VS content and a pH value of $28.75 \, g / L$, $18.90 \, g / L$ and $7.7 \, g / L$ respectively. The inoculum, $200 \, mL$, was added into an airtight $1000 \, mL$ transparent glass bottle (Simax, Czech Republic) and $100 \, mL$ of a glucose solution ($17.71 \, g / L$) was used as substrate. The final volume of $600 \, mL$ was reached by using tap water. 
 
 The reactors were connected to an automatic system for data acquisition and continuous pH control through NaOH addition as in \cite{DeGioannis2014}. Different pH set-point values in the range $5.5-7.0$ were adopted in the experiments. The substrate to inoculum ratio $F/M$ (Food/Microorganisms) was fixed at $1.5$ ($g$ COD\footnote{Where COD (Chemical Oxygen Demand) is a laboratory experimental measure of the organic compounds in a solid or liquid matrix.} substrate/$g$ VS inoculum).
 
 Each batch bioreactor was immersed into a thermostatic bath at $(35 \pm 1) \, {}^{\circ} C$ to ensure mesophilic conditions.
 This procedure was applied to three different DF reactors and, before closing the bottles, thirty minutes of nitrogen purge was carried out under anaerobic hood. The plastic cap of each reactor was equipped with sampling tubing. The liquid and gaseous samples were withdrawn by connecting the plastic cups to a gas measurement system (eudiometer) and a syringe, respectively.
 
 The total amount of the produced biogas was determined by water displacement. Biogas was characterized by a Varian Star 3400 gas chromatograph equipped with ShinCarbon ST 80/100 column and a thermal conductivity detector. Argon was used as carrier gas with $1.4 \, bar$ front and rear end pressure. The Organic Acids (OA) concentrations were quantified by high pressure liquid chromatography (HPLC) (Dionex LC 25 Chromatography Oven) equipped with a Synergi 4u Hydro RP 80A (size $250 \times 4.60 \, mm$) column and UV detector (Dionex AD25 Absorbance Detector). The cumulative normalized hydrogen production over time is reported in \figurename~\ref{f1.4}A.
     
\subsection{Photofermentative experiments}\label{n5.2}
	
 An adapted mixed culture of Purple Non-Sulphur Bacteria (PNSB) isolated from the Averno Lake (Naples, Italy) and enriched in a lab-scale reactor under continuous illumination was tested. In particular, the batch experiments were carried out in triplicate by using $500 \, mL$ borosilicate glass bottles GL 45 (Shott Duran, Germany) with a $400 \, mL$ working volume. The caps of each reactor were equipped with thin tubing on the top for sampling and gas extraction as in \cite{ghimire2016concomitant}. The light was continuously provided through fluorescent lamps which ensure constant illumination of $4000 \, lx$. The stirring conditions were fixed to $300 \, rpm$ through IKA RT 5 stirrer stations. The headspace of the reactors was flushed with argon gas for $20 \, min$ to avoid nitrogenase inhibition due to high nitrogen content \cite{trchounian2017improving}. The Photo Fermentation (PF) reactors were fed with a synthetic culture medium reproducing the characteristics of a real Dark Fermentation effluent \cite{luongo2017photofermentative}. The experiments were executed at room temperature ($25 \, {}^{\circ} C$).

 The sampling time for gas and liquid analysis varied from two to five days, depending on the PNSB growth phase and their activity. Hydrogen production was quantified through water displacement and gas chromatography by using a $1 \, L$ column fulfilled with an acidic solution ($1.5 \, \%$ HCl) to avoid biogas solubilisation. The hydrogen production was later calculated by considering the total biogas composition under normal conditions. The final hydrogen production was later normalized with respect to the organic load of glucose supplemented to the reactors (\figurename~\ref{f1.4}B).
 
\subsection{Microalgae Experiments}\label{n5.3}

 The experimental set-up was similar to the PF tests. The \textit{Chlorella pyrenoidosa} inoculum was provided by the ``Algae Biology Laboratory'' of the University of Naples ``\textit{Federico II''} (Italy). All the tests were carried out at room temperature in batch conditions by using $500 \, mL$ borosilicate glass bottles GL 45 (Shott Duran, Germany), constantly mixed at $200 \, rpm$. Continuous illumination was provided by using fluorescent lamps (average light intensity of $4000 \, lx$). 

 \begin{figure}[h]
 \begin{center}
 \includegraphics[width=.6\textwidth]{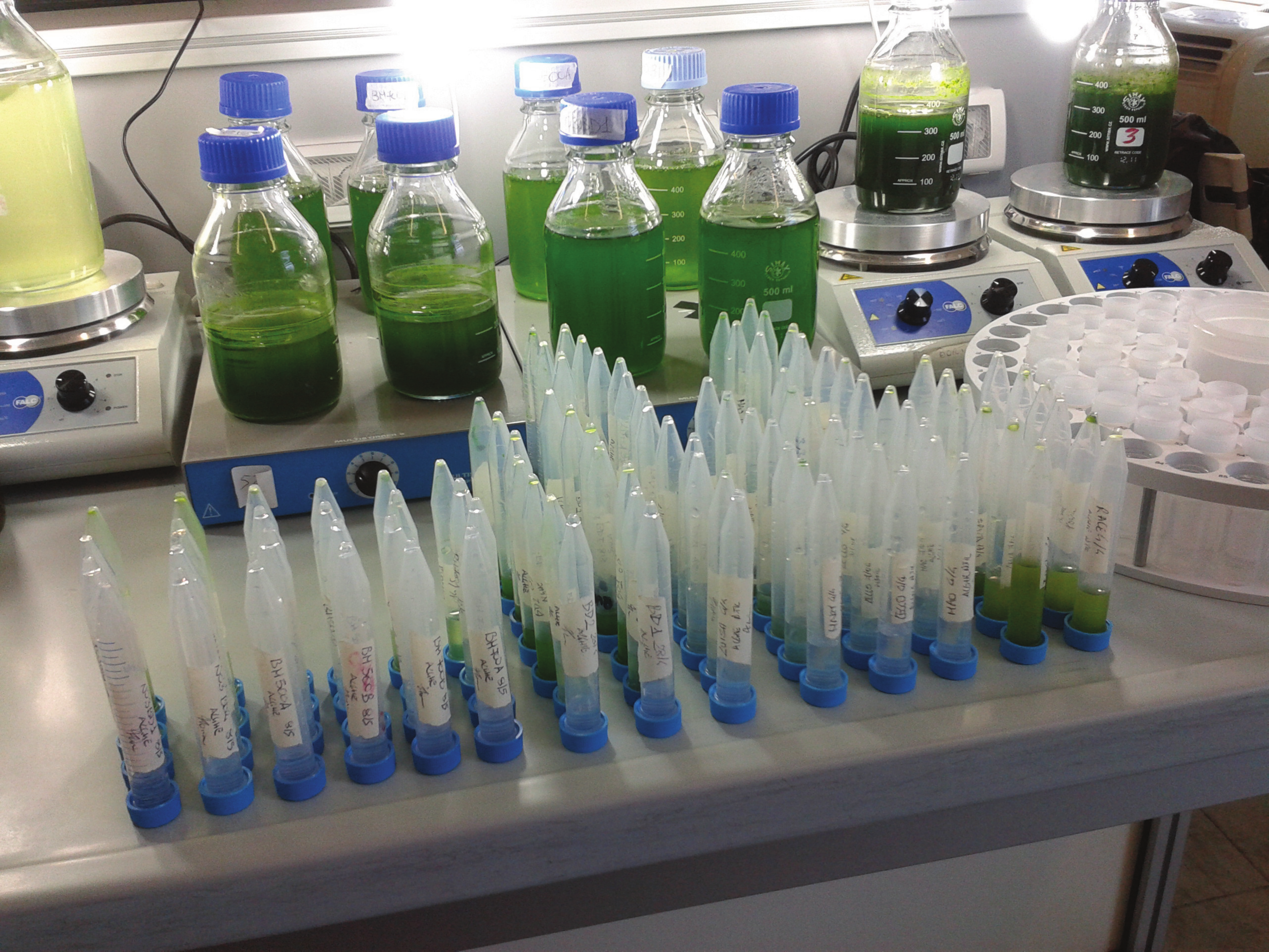}
 \caption{Algae experiment}
 \label{f1.3}
 \end{center}        
 \end{figure}
 
 A modified Bold's Basal Medium \cite{iasimone2017} with nitrate concentration around $180 \, mgNO_3 ^- / L$ was used for microalgae enrichment. Autotrophic growth was stimulated by the addition of inorganic carbon ($NaHCO_3$) until reach the $C/N$ (Carbon/Nitrogen) ratio of $15$. Biomass growth was monitored trough OD $680$ and a correlation curve was used to achieve the specific dry weight trend over time (\figurename~\ref{f1.4}C).

\section*{Acknowledgments}
This work has been carried out in the framework of the activities of the National Group of Mathematical Physics (GNFM, INdAM). Moreover, the work of L.F. and A.G. is partially supported by GNFM/INdAM Young Researchers Project 2017. Finally, L.F. and V.L. would also like to acknowledge the support of the project VOLAC - Valorization of OLive oil wastes for sustainable production of biocide-free Antibiofilm Compounds, funded by CARIPLO foundation.

	The experiments reported in this work have been performed by Dr Vincenzo Luongo at the Laboratory of Analysis and Environmental Research (LARA) of The Department of Civil, Architectural and Environmental Engineering (DICEA) of the University of Naples ``\emph{Federico II}''.
%
%
%
%
%%%%%%%BIBLIOGRAPHY
%
%
%
	
\end{document}